\documentclass[prb,twocolumn,showpacs,preprintnumbers,amsmath,amssymb]{revtex4}

\usepackage{graphicx}
\usepackage{dcolumn}
\usepackage{bm}

\begin{document}

\preprint{APS/123-QED}

\title{NMR study of the high--field magnetic phase of LiCuVO$_4$}

\author{N. B\"{u}ttgen$^1$\email{norbert.buettgen@physik.uni-augsburg.de}, W. Kraetschmer$^1$, L.E. Svistov$^2$, L.A.
Prozorova$^2$, and A. Prokofiev$^3$}

\affiliation{$^1$ Center for Electronic Correlations and Magnetism
EKM, Experimentalphysik V, Universit\"{a}t Augsburg, D--86135
Augsburg, Germany \\
$^2$ P.L.Kapitza Institute for Physical Problems RAS, 119334 Moscow, Russia\\
$^3$ Institut f\"{u}r Festk\"{o}rperphysik Technische
Universit\"{a}t Wien, A--1040 Wien, Austria}

\date{\today}

\begin{abstract}
We report on NMR studies of the quasi--1D antiferromagnetic $S=1/2$
chain cuprate LiCuVO$_4$, focusing on the high--field
spin--modulated phase observed recently in applied magnetic fields
$H
> H_{\rm c2}$ ($\mu_0H_{\rm c2} \approx 7.5$~T). The NMR spectra of
$^7$Li and $^{51}$V around the transition from the ordered to the
paramagnetic state were measured. It is shown that the
spin--modulated magnetic structure forms with ferromagnetic
interactions between spins of neighboring chains within the {\bf
ab}--plane  at low temperatures 0.6~K $ < T < T_{\rm N}$. The best
fit provides evidence that the mutual orientation between spins of
neighboring {\bf ab}--planes is random. For elevated temperatures
$T_{\rm N} < T \lesssim 15$~K, short--range magnetic order occurs at
least on the characteristic time scale of the NMR experiment.
\end{abstract}

\maketitle

\section{INTRODUCTION}
The problem of nontrivial ordering in frustrated quantum--spin
chains is considered theoretically as a challenging
issue.\cite{Chubukov_91,Kolezhuk_00,Kolezhuk_05,Dmitriev_08}
Recently, the quasi--one--dimensional (1D) antiferromagnetic $S=1/2$
chain cuprate LiCuVO$_4$ gained interest as a real material in this
context with a low magnetic ordering temperature $T_{\rm N}$ due to
weak inter--chain interactions.\cite{Sudan 2009,Heidrich 2009} In
this particular compound, magnetic frustration due to the
intra--chain nearest neighbor (NN) ferromagnetic and the
next--nearest neighbor (NNN) antiferromagnetic exchange interactions
yields an incommensurate helix structure of the magnetic Cu${}^{2+}$
moments. Additionally, the magnetic structure at $T_{\rm N}$ is
accompanied by ferroelectric order with spontaneous polarization at
the same temperature.\cite{Naito 07,Schrettle 08}

\begin{figure}
\includegraphics[width=60 mm,angle=-90,clip]{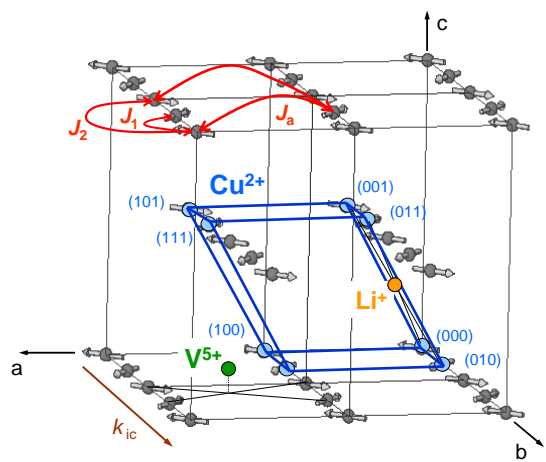}
\caption{Scheme of the Cu$^{2+}$ moments in the crystal structure of
LiCuVO$_4$ (Ref. \onlinecite{Gibson 04}). Copper ions are marked by
grey circles, where the arrows at each site constitute the helical
spin structure at $H=0$ below $T_{\rm N}$ (Ref. \onlinecite{Gibson
04}). $J_1,J_2,J_a$ (red arrows) denote the main exchange
integrals\cite{Enderle 05} defining the magnetic structure of
LiCuVO$_4$. Additionally, the positions of Li (orange) and V (green)
ions are exemplarily depicted by one ion each. In favor of clarity,
the V$^{5+}$ ion is shifted into the foreground of the cell.}
\label{Fig_1}
\end{figure}

LiCuVO$_4$ crystallizes in an inverse spinel structure $AB_2$O$_4$
with an orthorhombic distortion induced by a cooperative
Jahn--Teller effect of the Cu$^{2+}$ ions. The elementary cell
contains four magnetic ions Cu$^{2+}$($S=1/2$) with the coordinates
(0,0,0), (0,1/2,0), (1/2,0,1/2), and (1/2,1/2,1/2) (see Fig.
\ref{Fig_1}). From elastic neutron--diffraction experiments it was
established\cite{Gibson 04} that in the low temperature phase for $T
< T_N$ and zero applied magnetic field $H=0$ an incommensurate
planar spiral spin structure forms which has the propagation wave
vector $\mathbf{k_{ic}}$ directed along the Cu$^{2+}$ chains
($\mathbf{k_{ic}}\parallel\mathbf{b}$, see Fig. \ref{Fig_1}). We
parametrized the helix of this spin structure with magnetic moments
$\mathbf{\mu}$ of the Cu$^{2+}$ ions utilizing the coordinates $x$,
$y$, and $z$ along the $\mathbf{a}$,$\mathbf{b}$, and $\mathbf{c}$
directions, respectively (Ref. \onlinecite{Buettgen 07}):

\begin{eqnarray}\label{eq:1}
\mathbf{\mu}(x,y,z)=\mu_{\rm Cu} \cdot \mathbf{l_1} (-1)^{2z/c}
\cdot \cos(k_{ic}\cdot y+\phi)+
\\ \nonumber \mu_{\rm Cu} \cdot \mathbf{l_2} (-1)^{2z/c} \cdot \sin(k_{ic}\cdot
y+\phi),
\end{eqnarray}
where  $\mathbf{l_1}$ and $\mathbf{l_2}$ are orthogonal unit vectors
within the {\bf ab}--plane. At zero applied magnetic field $H=0$,
the absolute value of the propagation wave vector is k$_{\rm ic}$ =
$(1-0.532) \cdot 2\pi/b$ and the ordered Cu moment amounts to
$\mu_{\rm Cu}=0.31\mu_B$ (Refs. \onlinecite{Gibson 04,Enderle 05}).
The angle $\phi$ in Eq. \ref{eq:1} denotes an arbitrary phase shift.


Inelastic neutron--scattering experiments\cite{Enderle 05} confirmed
that the magnetically ordered structure of LiCuVO$_4$ is quasi--1D
with dominating exchange interactions $J_{1,2}$ within the chains.
It was shown that the incommensurate structure is due to strong
intra--chain NN ferromagnetic and NNN antiferromagnetic exchange
interactions with $J_1=-18$~K and $J_2=49$~K, respectively (see Fig.
\ref{Fig_1}). Note that this hierarchy of interactions causes strong
magnetic frustration. A three dimensional magnetic order results
from ferromagnetic exchange interactions $J_a \approx -4.3$~K
between magnetic moments of neighboring chains within an {\bf
ab}--plane (red arrows in Fig. \ref{Fig_1}), and about five times
weaker interactions between magnetic moments of different {\bf
ab}--planes.
\begin{figure}
\includegraphics[width=70 mm]{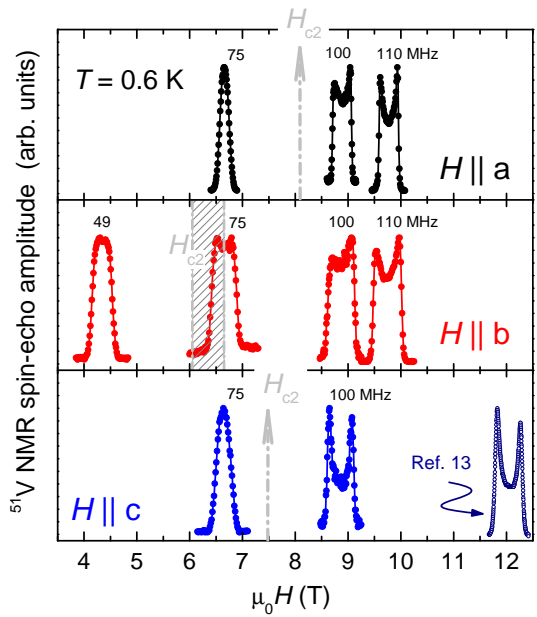}
\caption{NMR spectra of $^{51}$V measured at $T=0.6$~K for different
frequencies, i.e. different applied magnetic fields $H$. The
spectrum obtained in Ref. \onlinecite{Smith 06} at T = 1.64 K is
included.}
 \label{Fig_2}
\end{figure}
By application of a static magnetic field $H$, a number of
consecutive magnetic phase transitions was observed with increasing
field.\cite{Enderle 05, Buettgen 07, Banks 07} The saturation fields
were determined to be $\mu_0H_s \approx 40$~T and 50~T for the
orientations $\mathbf{H} \parallel \mathbf{c}$ and $\mathbf{H}
\parallel \mathbf{a}$, respectively. In the low--field range ($\mu_0H < 12$~T) the
magnetically ordered structures were studied by means of ESR and NMR
techniques.\cite{Buettgen 07} On increasing the field $\mu_0H > 0$,
the first phase transition takes place at $\mu_0H_{c1}\approx 2.5$~T
with the direction of $\mathbf{H}$ applied within the {\bf
ab}-plane. It can be explained as a spin--flop reorientation of the
spiral spin plane of the magnetic structure where the spin plane is
oriented perpendicularly to the applied magnetic field according to
$\mathbf{ H}\parallel (\mathbf{l_1}\times\mathbf{l_2})$. The
transition field $\mu_0H_{c1}$ is determined by the value of the
anisotropic exchange and by the anisotropy of the magnetic
susceptibility.

A more interesting and unexpected phase transition is observed at
higher fields $\mu_0H_{c2}\approx 7.5$~T. The observation of this
magnetic transition for all three directions $\mathbf{H}
\parallel \mathbf{a},\mathbf{b},\mathbf{c}$ reveals an exchange nature of
this transition. The NMR spectra observed at $H > H_{c2}$ can only
be explained by the assumption that a spin--modulated structure is
realized, where the ordered component of the spin structure is
oriented parallel to the applied magnetic field $\mathbf{H}
\parallel \mathbf{l_1}$ and $\mathbf{l_2}=0$. An indirect
confirmation of this sequence of magnetic phase transitions
suggested in Ref. \onlinecite{Buettgen 07} was obtained by
experimental investigation of the dielectric properties of
LiCuVO$_4$.\cite{Schrettle 08} A closer inspection of the
high--field phase for $H>H_{c2}$ will be given in this work by means
of NMR spectroscopy and different scenarios are discussed for the
collinear magnetic structure which is realized for $H>H_{c2}$.

\section{NMR EXPERIMENT}
The single crystal under investigation is the identical crystal used
in Refs. \onlinecite{Schrettle 08} and \onlinecite{Buettgen 07}. The
NMR experiments were performed with a phase coherent, homemade
spectrometer at radio frequencies within the range
$45.1<\nu<165$~MHz (for details see Ref. \onlinecite{Buettgen 07}).
The effective local magnetic field acting on nuclei of a nonmagnetic
ion is determined by long--range dipolar fields from the surrounding
magnetic ions and by so called 'contact' hyperfine fields from NN
magnetic ions. The positions of Li$^{1+}$ and V$^{5+}$ ions
surrounded by the copper ions are shown in figure \ref{Fig_1}. The
Li$^{1+}$ ions are located between the {\bf ab}--plane of Cu ions
and the V$^{5+}$ ions are very close to one single {\bf ab}--plane
of Cu ions. Therefore, the ${}^7$Li NMR spectra are sensitive to the
mutual orientation between Cu spins of adjacent {\bf ab}--planes. In
case of the ${}^{51}$V NMR in LiCuVO$_4$, this mutual orientation
between the Cu spins of adjacent {\bf ab}--planes does not affect
the spectral shape, because the effective local magnetic field at
the V sites is dominated by contact hyperfine fields from the four
Cu$^{+2}$ moments of the nearest {\bf ab}--plane.

\begin{figure}
\includegraphics[width=70 mm,angle=0]{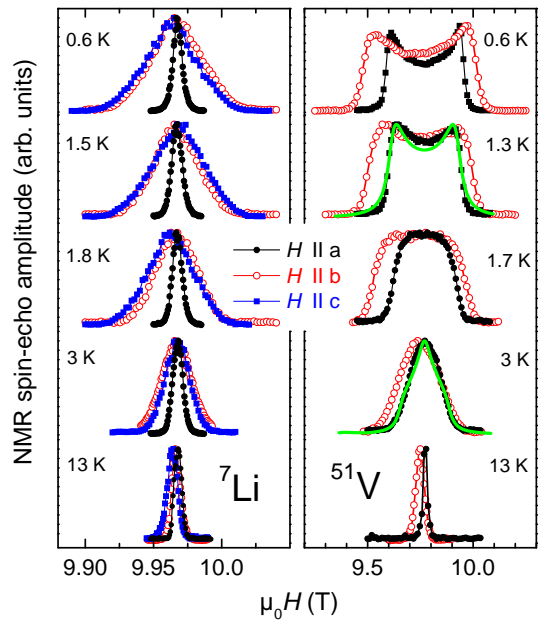}
\caption{ NMR spectra of $^{7}$Li (165 MHz, left column) and
$^{51}$V (110 MHz, right column) at different temperatures for
$H>H_{c2}$. Additionally, the $^{51}$V spectra at $T = 1.3$~K and
3~K for $\mathbf{H}\parallel\mathbf{a}$ are displayed together with
simulation curves (green solid lines).}
 \label{Fig_3}
\end{figure}
\begin{figure}[b]
\includegraphics[width=70 mm]{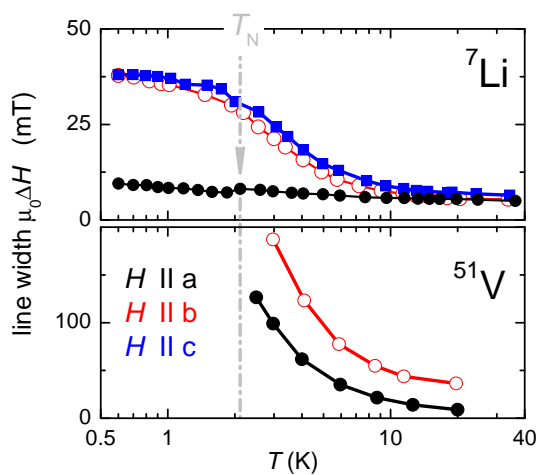}
\caption{Temperature dependences of the line widths $\Delta H(T)$
obtained from Gaussian fits to the data in Fig. \ref{Fig_3} at $H
> H_{\rm c2}$. Upper and lower frame for $^7$Li and $^{51}$V nuclei,
respectively. $T_{\rm N} \approx 2.1$~K is marked by an arrow (see
text).}
 \label{Fig_4}
\end{figure}

Figure \ref{Fig_2} shows the $^{51}$V NMR spectra at different
frequencies, i.e. different applied magnetic fields, for all three
directions $\mathbf{H} \parallel \mathbf{a,b,c}$. All spectra with
resonance fields $H<H_{c2}$ exhibit one single line, irrespective of
the orientation of the applied magnetic field. For fields $H >
H_{c2}$, the NMR spectra have the characteristic double--horn shape
which is a fingerprint of modulated magnetic structures. The arrows
in Fig. \ref{Fig_2} indicate the anisotropy of the transition fields
$H_{c2}$ obtained in Refs. \onlinecite{Schrettle 08,Buettgen 07} and
\onlinecite{Banks 07}. For $\mathbf{H}\parallel {\mathbf b}$, the
value for $H_{c2}$ is obtained from NMR data: at magnetic fields
$\mu_0H\approx 6.7$~T, the $^{51}$V NMR spectral line already
transforms to the double--horn pattern which is specific for the
high--field magnetic phase (see the middle frame in Fig.
\ref{Fig_2}). From the $^7$Li NMR spectral shape for $\mathbf{H}
\parallel {\mathbf b}$ it was concluded that $\mu_0H_{c2} > 6.05$~T
(cf. Ref. \onlinecite{Buettgen 07}). Therefore, we find the value of
$H_{c2}$ for the orientation $\mathbf{H} \parallel {\mathbf b}$
within the range $6.05 < \mu_0H_{c2} < 6.7$~T. This uncertainty is
depicted in the middle frame of Fig. \ref{Fig_2} by the hatched
area. The single line of the $^{51}$V NMR spectra which is obtained
for all three orientations for $H_{c1} < H < H_{c2}$ shows that the
spin rotating axis of the helical magnetic structure is parallel to
the applied field $\mathbf{H} \parallel
(\mathbf{l_1}\times\mathbf{l_2})$ confirming the observations made
in reference \onlinecite{Buettgen 07}.

Figure \ref{Fig_3} shows the temperature dependences of the NMR
spectra of $^7$Li and $^{51}$V at irradiation frequencies of 165~MHz
and 110~MHz, respectively. These frequencies were chosen in order to
obtain NMR spectra of both nuclei at the same applied magnetic field
around $\mu_0H = 10$~T, exceeding the transition field $H_{c2}$. The
$^7$Li spectra for all temperatures exhibit a single line whereas
the $^{51}$V spectra in the low--temperature range reveal the
double--horn pattern. For temperatures $T < 1.3$~K, the NMR spectral
shapes occur to be temperature independent. As it was shown in Ref.
\onlinecite{Buettgen 07} such spectral shapes of both nuclei can
only be realized for a collinear spin--modulated structure with
$\mathbf{H}\parallel \mathbf{l_1}$ and $\mathbf{l_2} = 0$. Figure
\ref{Fig_4} gives the temperature dependences of the NMR line width
of $^7$Li and $^{51}$V obtained from Gaussian fits of the data
presented in figure \ref{Fig_3}. In case of the $^{51}$V spectra,
such fitting to Gaussian lines was only possible for temperatures
$T>2.1$~K, because the spectral shape for $T < 2.1$~K changes to a
broad plateau--like pattern. It is important to note that there is
an abrupt 10 \% step--like decrease of the $^7$Li line width $\Delta
H(T)$ at around 2~K for the field orientation ${\mathbf H}
\parallel {\mathbf a}$. Thus, the temperature $T_{\rm N} \approx
2.1$~K identifies the transition temperature into the magnetically
ordered phase in agreement with the phase diagram established in
Refs. \onlinecite{Schrettle 08} and \onlinecite{Banks 07}. At
elevated temperatures $T_{\rm N} < T < 15$~K, the line widths
$\Delta H (T)$ for both nuclei decrease with increasing temperature.
A temperature independent, isotropic line width $\Delta H$ which is
characteristic for paramagnetic behavior is only established for
$T>15$~K, far above $T_{\rm N}$.

\section{Discussion and Conclusion}
In order to further investigate this high--field phase $H
> H_{c2}$ we will consider two types of magnetic structures
which assure such modulated spin components directed along $H$, i.e.
a planar spiral spin structure with $\mathbf{H} \perp \mathbf{n}$
(where $\mathbf{n} = \mathbf{l_1\times l_2}$) and a collinear
structure with $\mathbf{H} \parallel \mathbf{l_1}$, respectively.

Moreover, we take into account the hierarchy of the magnetic
exchange interactions in LiCuVO$_4$. As the exchange interactions
within the {\bf ab}--plane greatly exceed the inter--plane
interactions, we will test on one hand the case with long--range
magnetic order in {\bf c}--direction (3D in the following), and on
the other hand the case of a long--range order within the {\bf
ab}--planes but an arbitrary mutual orientation of spins along the
{\bf c}--direction, i.e. between different {\bf ab}--planes (2D).

\begin{figure}
\includegraphics[width=70 mm]{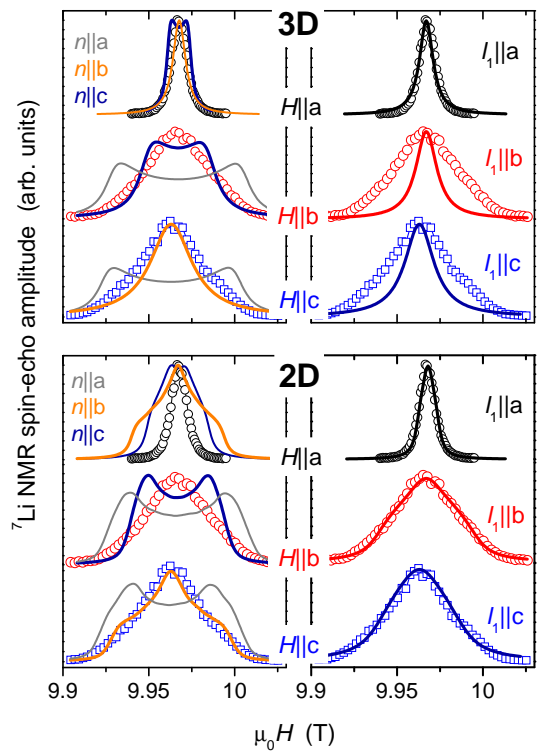}
\caption{$^7$Li NMR spectra at $T=0.6$~K and 165~MHz (symbols). The
solid lines represent the simulations (see text): left (right)
column shows the simulated spectra for the planar spiral structure
with $\mathbf{H} \perp \mathbf{n}$ (collinear spin--modulated
structure, $\mathbf{H} \parallel \mathbf{l_1}$ and $\mathbf{l_2}=0$
), respectively.}
 \label{Fig_5}
\end{figure}

Figure \ref{Fig_5} presents the experimental $^7$Li NMR spectra at
$T = 0.6$~K (symbols) and the simulated spectra (solid lines)
according to the suggested spin structures mentioned above. The
simulations result from calculations of the Cu$^{2+}$ dipolar fields
at the probing $^7$Li nuclear site (cf. Ref. \onlinecite{Buettgen
07}). The left (right) column of Fig. \ref{Fig_5} shows the
simulated $^7$Li NMR spectra for the planar spiral spin structure
(collinear spin--modulated structure), respectively. The 3D
long--range magnetic order is taken into account by a constant phase
$\phi(x,z)$ in Eq. \ref{eq:1}, whereas in case of the 2D long--range
magnetic order the phase $\phi$ is only constant with respect to $x$
but arbitrary with respect to the $z$--coordinate
[$\phi(x,z)=random(z)$]. The best agreement between experiment and
simulations is obtained for the spin--modulated 2D magnetic
structure, i.e. with a random phase relation between spins along the
$c$--axis (cf. lower right frame in Fig. \ref{Fig_5}).

The incommensurate propagation wave vector $\mathbf{k_{\rm ic}}$ at
high applied magnetic fields is experimentally unknown, but the
simulations revealed that a variation of $k_{\rm ic}$ does not
change the shape of the calculated NMR spectra. Thus, at low
temperatures and high magnetic fields $H > H_{c2}$ long--range
magnetic order in the {\bf ab}--plane without inter--plane coupling
along the $c$--direction is realized in LiCuVO$_4$. This proposed
magnetic structure is consistent with the $^{51}$V NMR spectra: the
effective magnetic fields on these nuclei are dominated by contact
fields of the four neighboring Cu ions of the nearest {\bf
ab}--plane (see Fig. \ref{Fig_1}). Therefore, these spectra are not
sensitive to magnetic disorder in $c$--direction. This explains the
double--horn shape of the $^{51}$V NMR spectra for $H > H_{c2}$ at
very low temperatures $T=0.6$~K. In addition, the lack of a spiral
spin structure proposed for $H > H_{c2}$ accounts for the
paraelectric behavior according to reference \onlinecite{Schrettle
08}.

At elevated temperatures $T \gtrsim 2$~K, the shape of the $^{51}$V
NMR spectra transforms from the double--horn shape at low
temperatures to a single line pattern which can be fitted by
Gaussian lines. In the lower frame of figure \ref{Fig_4}, the
temperature dependence of the line width $\Delta H$ of these
Gaussian lines is plotted. It is important to note that broad
Gaussian lines can be observed up to approximately 15~K indicating
short--range magnetic ordering at the time scale of the NMR
experiment, far above the ordering temperature $T_{\rm N} \approx
2$~K. A lower bound of the time scale is given by the time for
spin--echo formation which was achieved in our experiments with
pulse separation times $\tau_D = 35\mu {\rm s}$. From the
spin--lattice relaxation time $T_1$ of our measurements we estimate
an extension of the time scale even to milliseconds.

Most probably this transformation of the $^{51}$V NMR spectral shape
(see right column in Fig. \ref{Fig_3}) indicates the loss of
ferromagnetic interactions along the $\mathbf{a}$--direction towards
higher temperatures. In order to corroborate this assumption we
simulated the $^{51}$V NMR spectra at 1.3 and 3~K for this scenario.
The simulated spectra are included in Fig. \ref{Fig_3} (right
column, green solid lines): the simulation at 1.3~K is based on the
spin--modulated 2D magnetic structure with random phase relation
$\phi$ along the {\bf c}--direction mentioned above for the $^{7}$Li
NMR spectra. For the simulation at 3~K, the release of ferromagnetic
interactions between neighboring chains in the
$\mathbf{a}$--direction towards higher temperatures is taken into
account by randomizing the phase relation $\phi$ along the {\bf
a}--direction as well [$\phi(x,z) = random(x,z)$ in Eq. \ref{eq:1}].

In conclusion, the magnetic structure of the high--field magnetic
phase of the quasi--1D antiferromagnet LiCuVO$_4$ was studied by NMR
experiments. We determined that the spin--modulated magnetic
structure ($\mathbf{H} \parallel \mathbf{l_1}$) with long--range
magnetic order within the {\bf ab}--plane and a random phase
relation between the spins of neighboring {\bf ab}--planes is
realized in LiCuVO$_4$ at $H > H_{\rm c2}$ and low temperatures $T <
T_{\rm N}$. The observed NMR spectra can be satisfactorily described
by the following structure:
\begin{eqnarray}\label{eq:2}
\mathbf{\mu}(x,y,z)=\mu_{\rm Cu} \cdot \mathbf{l}\cdot
\cos[k_{ic}\cdot y+\phi(z)],
\end{eqnarray}
where $\mathbf{l}$ is the unit vector parallel to the applied
magnetic field $\mathbf{H}$ and the phase $\phi(z)$ between adjacent
spins in {\bf c}--direction is random. Within the temperature range
$T_{\rm N} < T \lesssim 15$~K, the NMR spectra can be described if
we assume that the phase relation $\phi$ between neighboring spin
chains in the {\bf ab}--plane is random too. The particular phases
for both structures below and above $T_{\rm N}$ for $H > H_{\rm c2}$
are time independent at least on the characteristic time scale of
the NMR experiment of microseconds.

\begin{acknowledgments}
This work is supported by the Grants 09- 0212341, 10-0201105-a
Russian Foundation for Basic Research, Russian President Program of
Scientific Schools, and by the German Research Society (DFG) within
the Transregional Collaborative Research Center (TRR 80).
\end{acknowledgments}

\end{document}